\RequirePackage{tikz} 
\usetikzlibrary{shapes,arrows,automata}
\documentclass[fleqn,10pt]{wlscirep}
\usepackage[utf8]{inputenc}
\usepackage[T1]{fontenc}
\usepackage{tikz}
\usetikzlibrary{shapes,arrows}
\usepackage{dot2texi}
\usepackage{graphicx}
\usepackage{float}
\usepackage{caption}
\usepackage{subcaption}

\long\def\omitit#1{}
\title{Analysis of the competition among viral strains using a temporal interaction-driven contagion model}

\author[1]{Alex Abbey}
\author[2]{Yuval Shahar}
\author[1,*]{Osnat Mokryn}
\affil[1]{Information Systems, University of Haifa, Haifa, Israel}
\affil[2]{Software and Information Systems Engineering, Ben Gurion University, Beer Sheva, Israel}

\affil[*]{corresponding.ossimo@gmail.com}


\begin{abstract}
The temporal dynamics of social interactions were shown to influence the spread of disease. Here, we model the conditions of progression and competition for several viral strains, exploring various levels of cross-immunity  over temporal networks. We use our interaction-driven contagion model and characterize, using it, several viral variants. Our results, obtained on temporal random networks and on real-world interaction data, demonstrate that temporal dynamics are crucial to determining the competition results. We consider two and three competing pathogens and show the conditions under which a \textit{slower} pathogen will remain active and create a second wave infecting most of the population. We then show that when the \textit{duration} of the encounters is considered, the spreading dynamics change significantly. Our results indicate that when considering airborne diseases, it might be crucial to consider the duration of temporal meetings to model the spread of pathogens in a population.
\end{abstract}
\begin{document}

\flushbottom
\maketitle
\thispagestyle{empty}

\section*{Introduction}
Human communities are complex social systems in which emergent properties - such as the dynamic nature of a viral infection - result from their members' local, uncoordinated temporal interactions \cite{eubank2004modelling,pastor2015epidemic,masuda2017introduction}. 
Viral transmission is mediated through the community's members' temporal interactions in the community. The sequence and order of these interactions, termed time-respecting paths, are critical for accurate modeling of the progression of a disease~\cite{holme2012temporal,ENRIGHT201888,walensky2021sars}.

Here, we use temporal networks with time-respecting paths and continuous-time network histories~\cite{zhang2017random,abbey2022interactionbased} as well as a real-world contact network~\cite{sapiezynski2019interaction} to model multiple competing variants in a population. As SARS-CoV-2 continues to mutate, some of its variants become variants of concern. These variants either differ in their transmissibility, infectivity,  pathogenicity or their ability to evade immunity~\cite{moore2021sars,harvey2021sars,alpert2021early,mahase2021covid}. 

We quantify the competition conditions between variants that differ in their transmissibility and under changing levels of cross-immunity, over temporal random networks, with our interaction-driven contagion model~\cite{abbey2022interactionbased}. Previous studies considered two competing pathogens in a population with full cross-immunity. Being infected by one of the pathogens provides subsequent immunity to both. Studies have demonstrated that the faster pathogen is more likely to dominate the network~\cite{karrer2011competing,wang2019coevolution,mann2021two,okabe2022spread}. When different cross-immunity levels are considered, this is not necessarily the case~\cite{poletto2015characterising}. However, these studies did not consider the temporal ordering of events and the time-respecting paths.

There are numerous practical considerations when considering competition conditions between variants in a population. Co-existing variants compete on a shared resource that is the transmission vehicle, i.e., the population~\cite{bono2013competition}. There is an interplay between the network's dynamics and the dynamics of the disease that determines the course of the disease~\cite{sayama2015introduction,liu2018measurability}. For example, a variant that leads to more severe disease in the population causes more people to become more severely ill and for a longer duration~\cite{vafadar2021competitive}, thus changing the network's dynamics as people are removed from the network for longer periods than is the case with other variants. Variants may also differ in their initial conditions. They may arrive at different stages of immunity or differ in their transmission opportunities~\cite{walensky2021sars,ALPERT20212595,alpert2021early,twohig2022hospital}.  

We consider competing variants under changing conditions of cross-immunity over two types of temporal networks: temporal random networks and real-life contacts, using our interaction-driven SEIR-like model for temporal networks~\cite{abbey2022interactionbased}. 
The advantages of working with real-life temporal interactions are numerous, as  human interactions are bursty, temporal, highly contextual and networked~\cite{barabasi2005origin,holme2012temporal,mokryn2016role,fumanelli2012inferring,pastor2015epidemic,delvenne2015diffusion}.   
We model the competition between multiple variants on temporal networks while considering various cross-immunity dynamics. 

Our surprising results are that slower variants continue to infect in the presence of faster (and even much faster) variants. When there is no cross-immunity, the slower variant creates a second wave and reinfects the population previously infected by the faster variant. We further determine, using a heatmap of total infection ratio between the variants under changing conditions of cross-immunity, the conditions in which a slow variant would be the dominant variant in a second phase and will infect the majority of the population.  

We then proceed to model the competition between three variants in random temporal networks and real-world contact networks, with either full cross-immunity or no cross-immunity at all, with our SEIR-like interaction-driven model. We show that both networks provide similar results for the competition under full cross-immunity when considering maximal infection probabilities. When there is no cross-immunity, the dynamics of all three variants are similar over the real-world network, with the faster variants, i.e., with a higher probability of infection, infecting a larger part of the population than the slower ones.  %rate that is  and there is almost a linear ratio between the infection rate and the probability of infection. However, the contagious process in random and real-world networks differ when reinfection with other variants is possible. In the temporal random networks all three variants infect parts of the population, somewhat corresponding to the ratio between them. However, in the real-world temporal network, when three variants compete, the slowest variant dies out, while the two others compete. 

The competition dynamics differ substantially when the duration of the meetings is taken as a factor in the contagious process. When variants differ by the minimal time it takes them to infect, with faster variants taking less time to infect than slower variants, the competing conditions become more complex. The slowest variant dies out, leaving the two faster ones to compete.  The fast variant dominates a larger part of the network and for a longer duration, followed by the second fastest variant, which creates a large second wave in which it infects at a faster rate than the rate at which that variant infected during the first wave.

Thus, we show here that when the duration of meetings is considered in the modeling of temporal networks, the spreading process of competing variants is substantially different.

\section*{Results}
\subsection*{Race conditions between two variants with changing values of cross-immunity}
\begin{figure}[!ht] 
  \begin{subfigure}[b]{0.5\linewidth}
    \centering
    \includegraphics[width=0.99\linewidth]{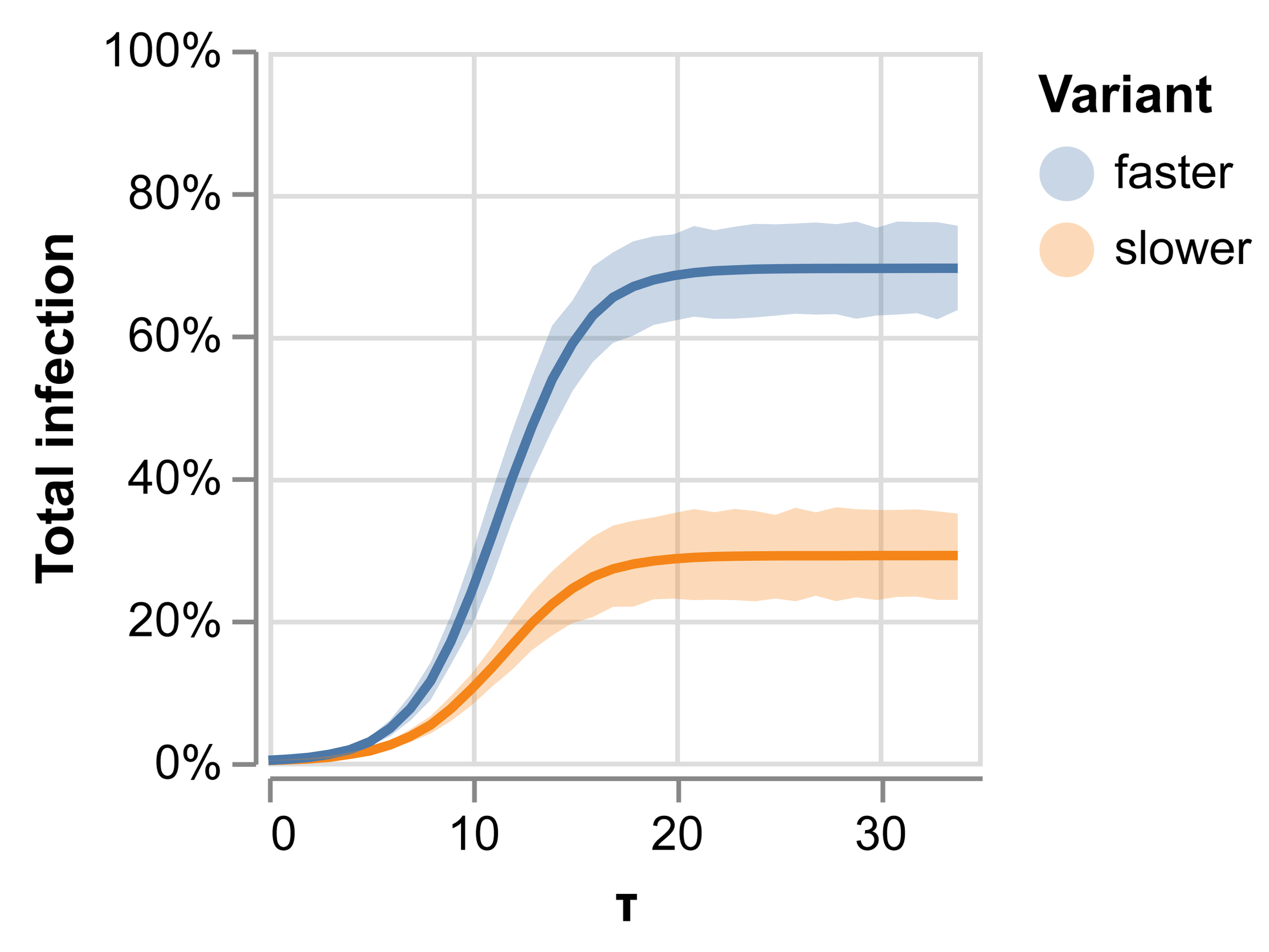} 
    \caption{Full cross-immunity} 
    \label{fig:sa} 
    %\vspace{4ex}
  \end{subfigure}%% 
  \begin{subfigure}[b]{0.5\linewidth}
    \centering
    \includegraphics[width=0.99\linewidth]{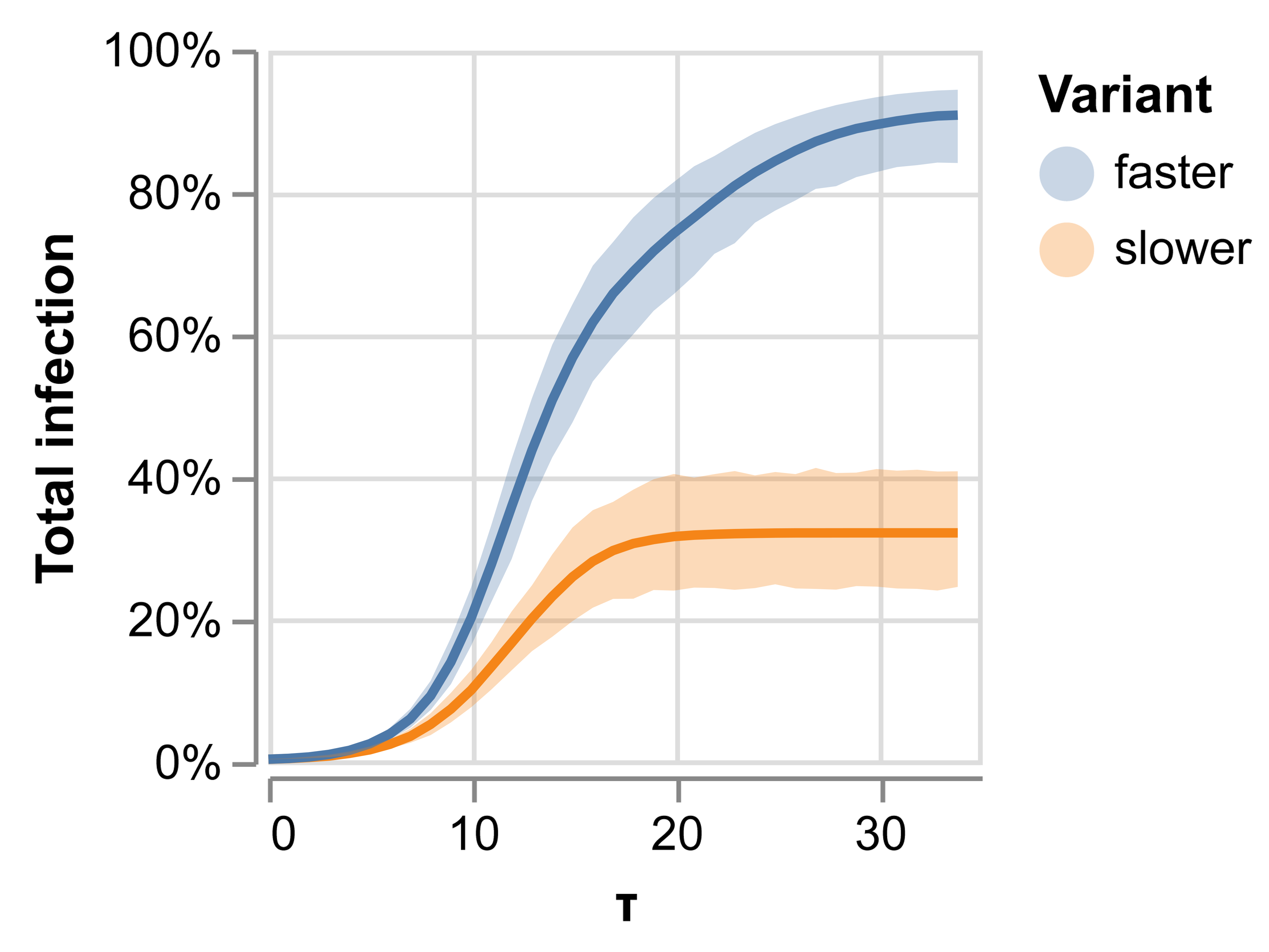} 
    \caption{Cross-immunity from faster to slower} 
    \label{fig:sb} 
    %\vspace{4ex}
  \end{subfigure} 
  \begin{subfigure}[b]{0.5\linewidth}
    \centering
    \includegraphics[width=0.99\linewidth]{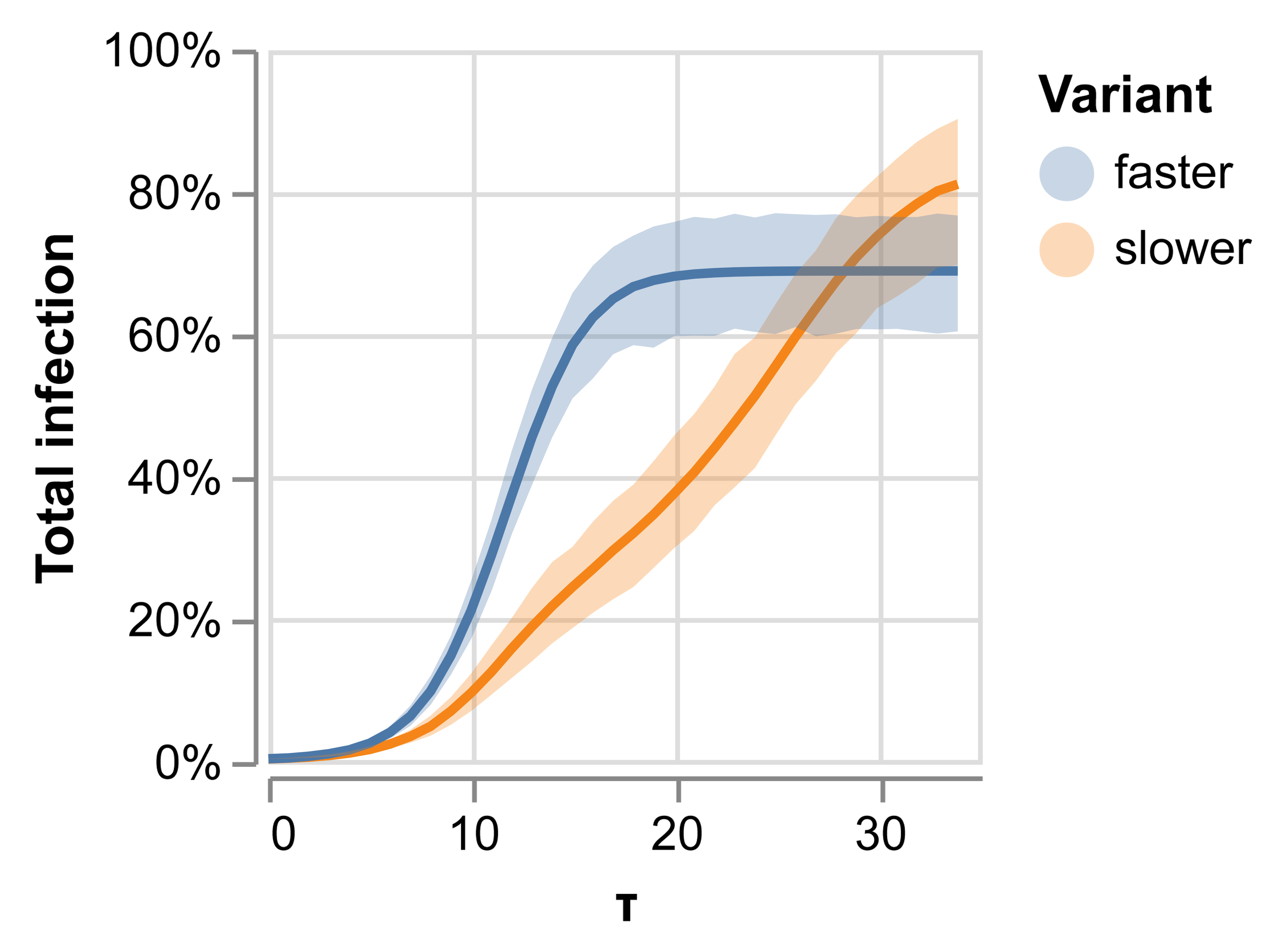} 
    \caption{Cross-immunity from slower to faster} 
    \label{fig:sc} 
  \end{subfigure}%%
  \begin{subfigure}[b]{0.5\linewidth}
    \centering
    \includegraphics[width=0.99\linewidth]{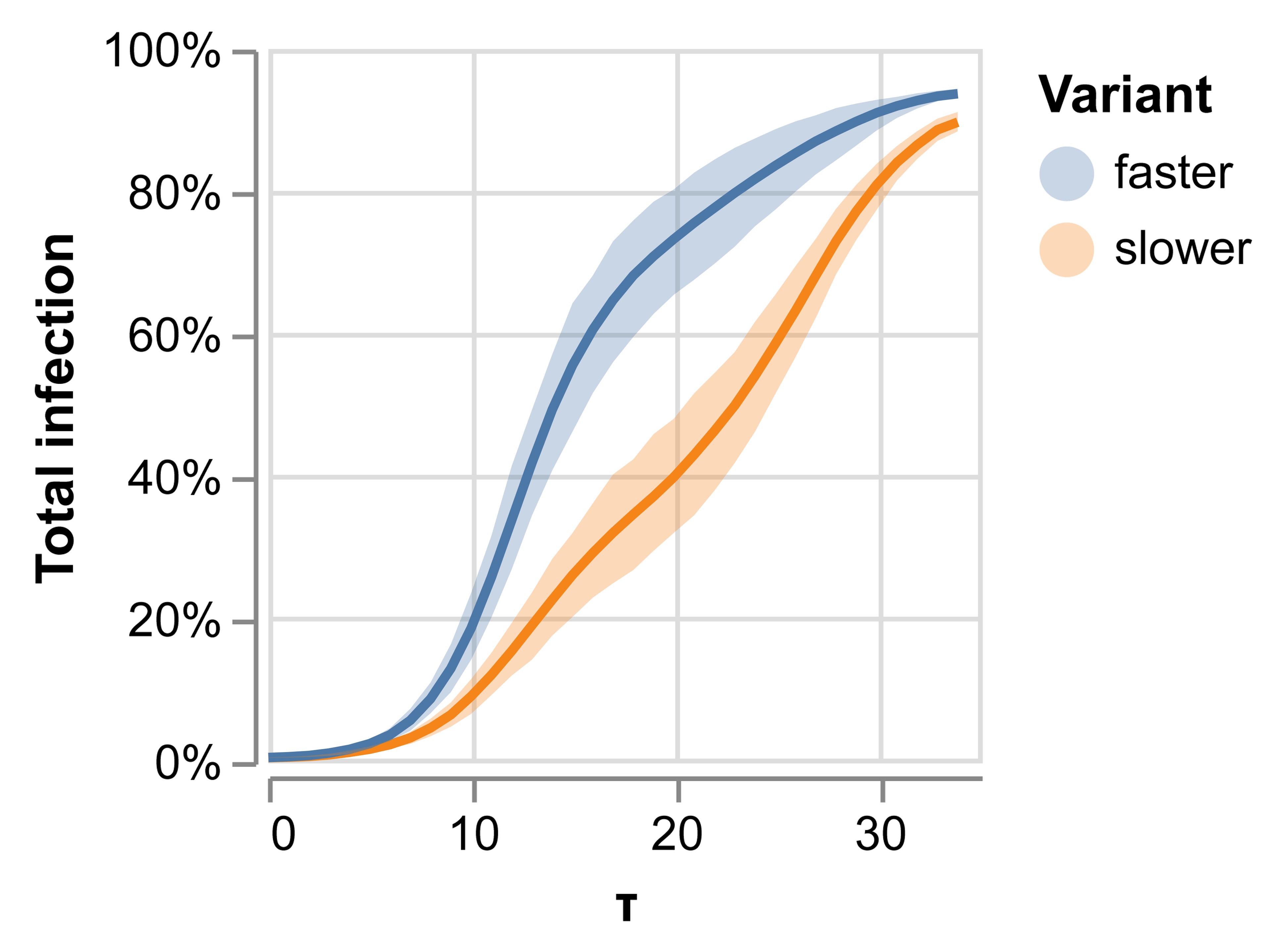} 
    \caption{No immunity to either} 
    \label{fig:sd} 
  \end{subfigure} 
  \caption{The aggregated percentage of infections in the population for each of the variants under (a) full cross-immunity (no reinfection in either), (b) recovery from the slower variant does not provide immunity to the faster variant, (c) recovery from the faster variant does not provide immunity to the slower variant, (d) no immunity to either. }
  \label{fig:4edges} 
\end{figure}

We created a temporal random network with $10,000$ time windows of $1000$ nodes each using the RDG package~\cite{abbey2022interactionbased}. A time window duration is denoted by $\tau$, and consists of $288 \times 5\text{-minute}$ intervals. Two variants were determined, a slow variant, and a fast variant, s.t. \[P_{\text{max}}^{\text{ fast}} = 1.2 \cdot P_{\text{max}}^{\text{ slow}}\]
Where $P_{\text{max}}^{\text{ fast}},P_{\text{max}}^{\text{ slow}}$ are the probabilities of being infected in maximum exposure to the fast (or slow) variant, as defined in Equation~\ref{eq:exp}. The values were chosen to allow for competition conditions. Both variants begin simultaneously with a randomly chosen patient zero in the first time window. Each experiment was repeated 200 times. We use our SEIR-like model.
Figure~\ref{fig:4edges} describes the results of our first experiment. Here, we examine the competition under various immunity conditions. In Figure~\ref{fig:sa} we can see that when there is full cross-immunity, the faster variant infects the majority of the population, while the slower one obtains much less. Both variants plateau at the same time. This changes when a breakthrough infection is possible with the faster variant, as can be seen in Figure~\ref{fig:sb}. Here, the slower variant infects a small portion of the population, who, once recovered, infected again with the faster variant. When the breakthrough probability is from the faster variant to the slower one, i.e., cross-immunity only from the slow variant to the fast one, we can see in Figure~\ref{fig:sc} that the fast variant saturates when it infects around 60\% of the population. Then the slower variant takes over and infects the vast majority of the population, infecting close to 90\% of the population. In Figure~\ref{fig:sd} we see that when both variants can reinfect, most of the population would first be infected by the fast variant but then, after recovery, would also be infected with the slower one. 

\begin{figure}[th]%
\centering
\includegraphics[width=\textwidth]{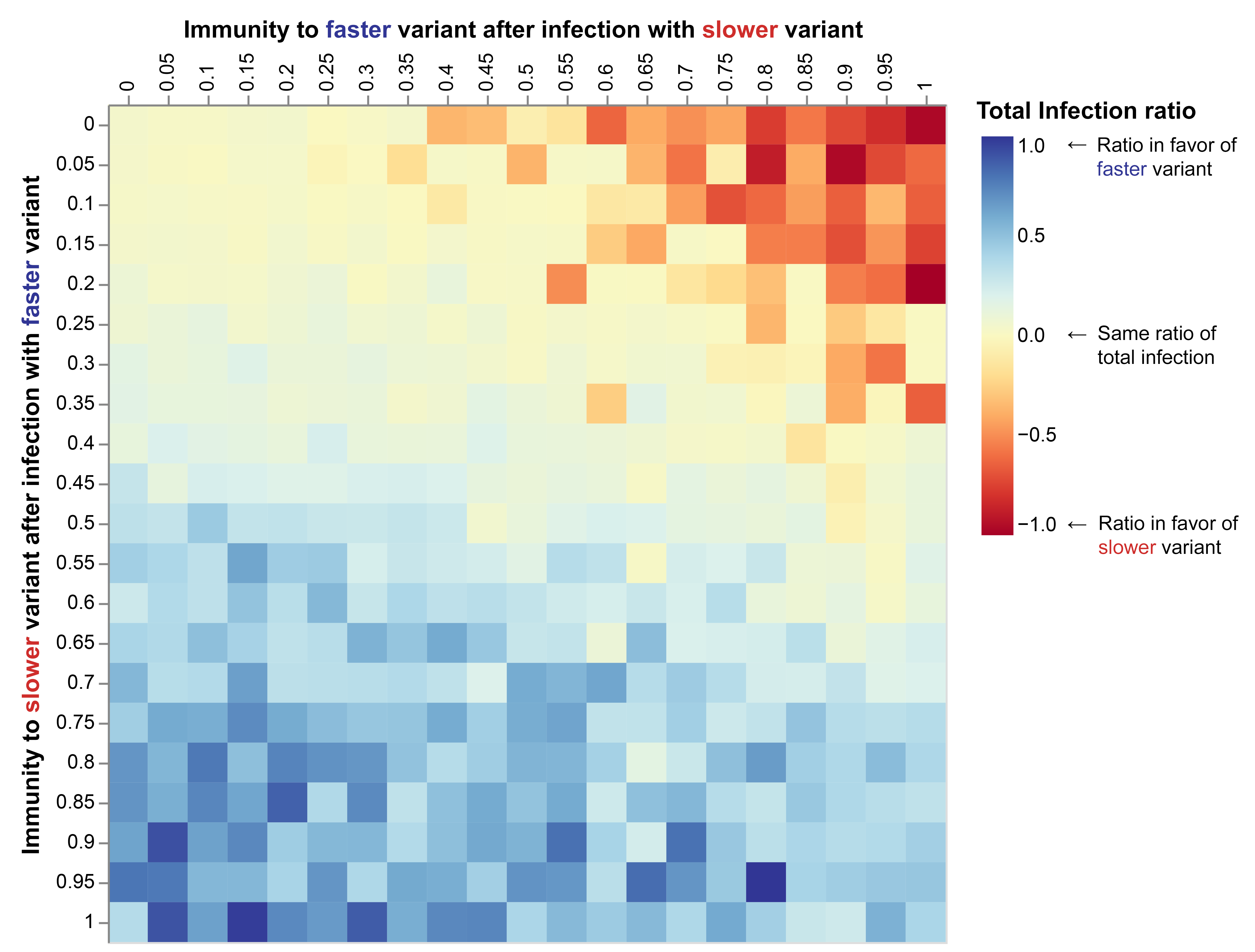}
\caption{A heatmap of the total infection ratio between the fast and the slow variants under changing conditions of cross-immunity. Immunity between the variants changes from no immunity to full immunity,  that is $\{\chi_{\text{fast} \rightarrow \text{slow}}, \chi_{\text{slow} \rightarrow \text{fast}}\} \in [0,0.05, \ldots 1]$. In the heatmap, low values (denoted in red) correspond to the slow variant infecting the majority of the population, high values (blue) to the faster variant infecting the majority.}\label{figrc}
\end{figure}

We continue to evaluate the infection probability of each of the variants, slower and faster, under changing conditions of cross-immunity. Figure~\ref{figrc} depicts the total infection ratio between the fast and the slower variants. Let us term the fast variant as $f$ and the slow variant $s$. The cross-immunity probability was changed in steps of $0.05$ from zero to one for each of the variants ($\{\chi_{{f} \rightarrow {s}}, \chi_{{s} \rightarrow {f}}\} \in [0,0.05, \ldots 1]$). Each of these combinations was iterated 30 times, to a total of $12,000$ iterations for the heatmap. The results were normalized to the range $[-1,1]$. The heatmap shows that in the majority of the cases, the fast variant is the dominant variant, and the ratio is more on the blue scale. When there is no cross-immunity between the variants, i.e., a person can be reinfected with a variant after recovering from the other, both variants manage to reinfect the entire population. This corresponds to the upper left corner of the heatmap. In figure~\ref{fig:sd} we see the aggregated infection in the population per variant with $\chi_{{f} \rightarrow {s}} = \chi_{{s} \rightarrow {f}} = 0$. Both variants infect the population at a different rate. When the faster variant saturates, the slower variant reinfects those that recovered from the fast variant in a second wave. Thus, the fast variant infects the population in a first wave, and the slower variant, infecting some of the population initially, infects the rest in a second wave. From Figure~\ref{figrc} we  see that this scenario describes the area in which  $\chi_{{f} \rightarrow {s}} \leq 0.4$ and $\chi_{{s} \rightarrow {f}} \leq 0.3$.  

However, interestingly, the slow variant dominates the population, and the ratio calculated in the heatmap (the red zone in the upper right corner of the heatmap in Figure~\ref{figrc}) greatly favors the slow strain, when infection with the slower variant results in immunity to the faster one but not vice-versa. In our experiments, the conditions are that $\chi_{{s} \rightarrow {f}} \geq 0.5$ and  $\chi_{{f} \rightarrow {s}} \leq 0.4$.

\subsection*{Conditions that allow competition}
It is then left to see under which conditions a slower variant can create a second wave of infection. %Okabe and Shudo~\cite{okabe2022spread} determined that for two similarly contagious variants, the faster one will not dominate under full cross-immunity. We continue here to examine the competition when there is no cross-immunity to either.  
\begin{figure}[h]%
\centering
\includegraphics[width=\textwidth]{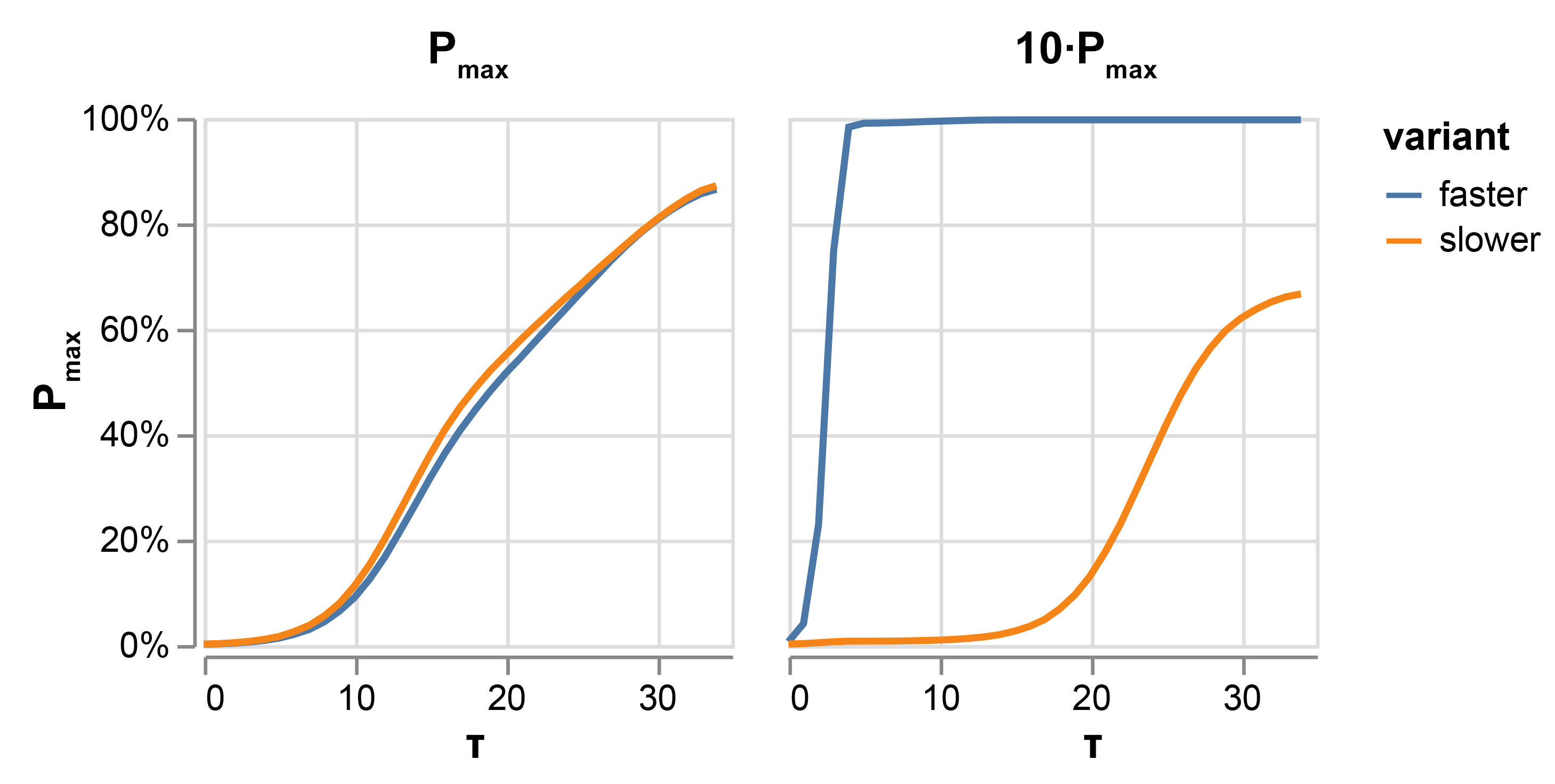}
\caption{Determining competition conditions: The aggregated infection rate of each variant over a temporal random network. Left panel: two variants with the same infection probability.  Right panel: the blue variant is ten times more infectious.}\label{fig:p1p2}
\end{figure}
Figure~\ref{fig:p1p2} shows the aggregated percentage of infections in a population for two competing variants. The left panel depicts two similar variants. Both variants advance at the same rate and can reinfect the population recovered from the competing variant. The right panel shows two variants, with the faster one ten times more contagious than the slower one. The fast variant infects all of the population rapidly. Interestingly, the much slower variant does not die out and keeps infecting at a low rate,  creating a second wave in which it slowly reinfects the population that has recovered from the fast variant. We find that when reinfection with the competing variant is possible, slower variants can create a second wave of infection.

%Note that different Pmax values may lead to different outcomes for different network's density, as is further explained in (Arxiv). Here, we chose Pmax values such that allow for changing race conditions, as described ahead. 
\begin{figure}[h]%
\centering
\includegraphics[width=\textwidth]{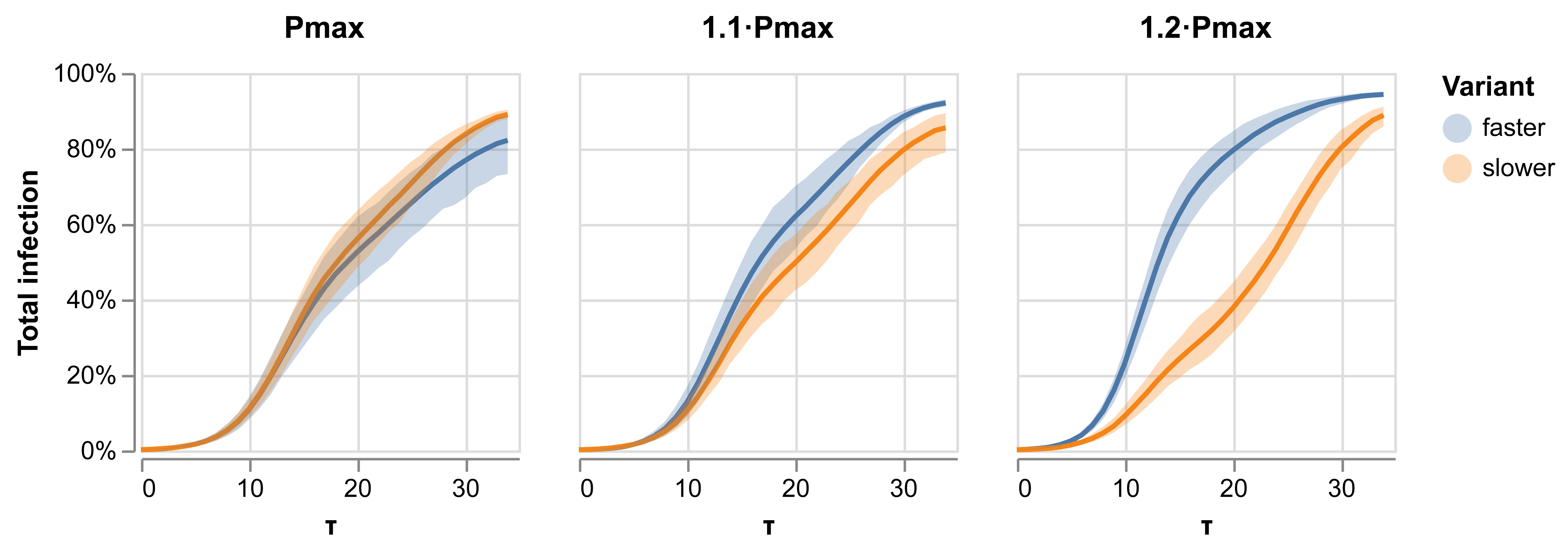}
\caption{Competition between variants with no cross-immunity. Left panel: similar variants. Middle panel: faster variant is 1.1 times more probable to infect. Right panel: faster variant is 1.2 times more probable to infect. }\label{figp-choice}
\end{figure}
Figure~\ref{figp-choice}  depicts the competition between variants with no cross-immunity for variants that are either similar (on the left panel) or differ in their probability of infection. In the middle panel, the variant is 1.1 times more likely to infect, and in the right panel, 1.2 times. We then chose the competing conditions such that the faster variant is 1.2 times more probable to infect than the slower one. This allows for a faster and a slower variant.% This allows for a more competition while one of the variants infects at a different rate. %Similar to what was previously found, the infection rate of the faster variant is non-linear~\cite{okabe2022spread}. 

\subsection*{Three competing variants}

\begin{figure}[!ht] 
  \begin{subfigure}[b]{0.5\linewidth}
    \centering
   \includegraphics[width=0.85\linewidth]{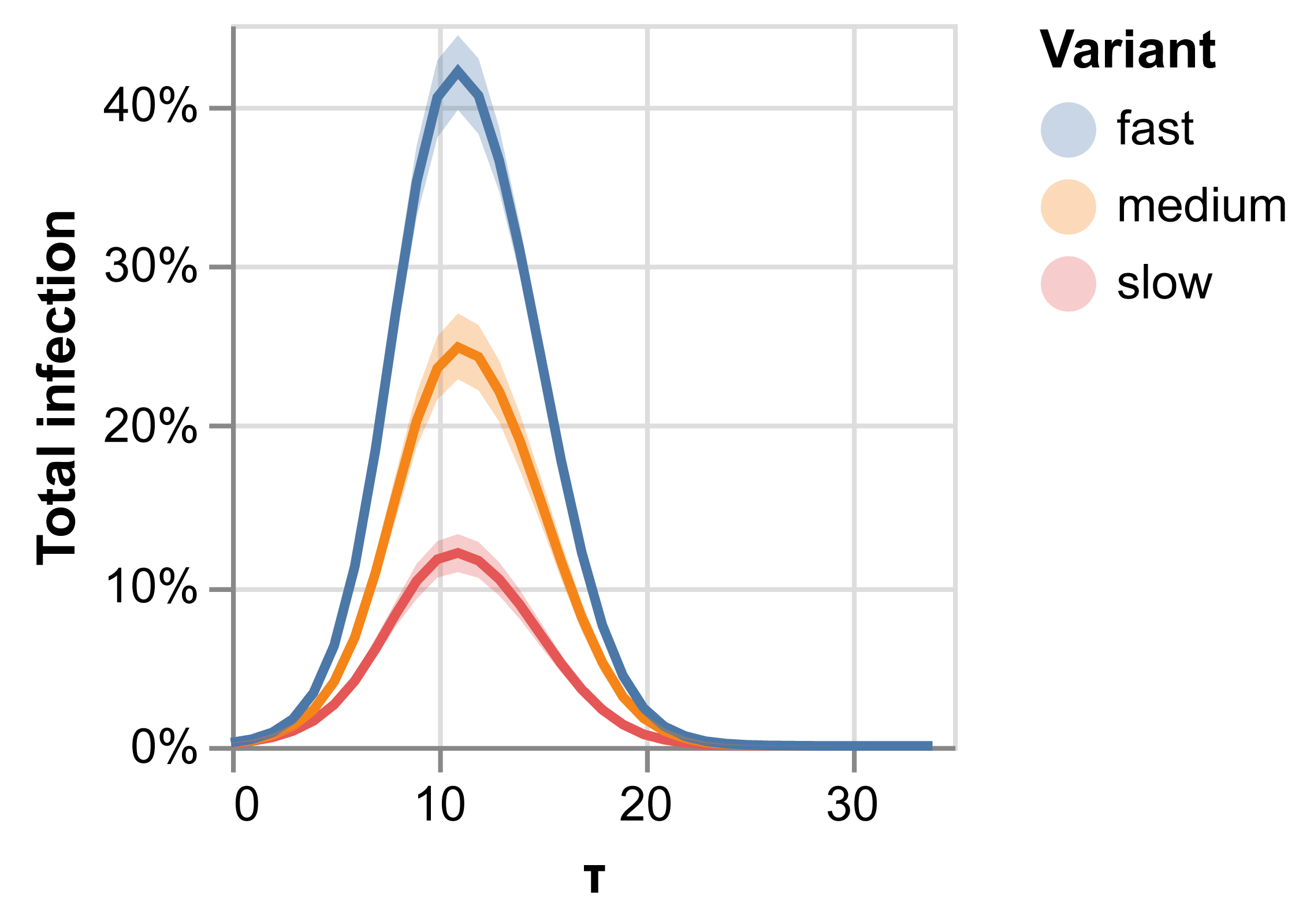} 
    \caption{Cross-immunity aggregated daily infections} 
    \label{fig:3a} 
    %\vspace{4ex}
  \end{subfigure}%% 
  \begin{subfigure}[b]{0.5\linewidth}
    \centering
     \includegraphics[width=0.85\linewidth]{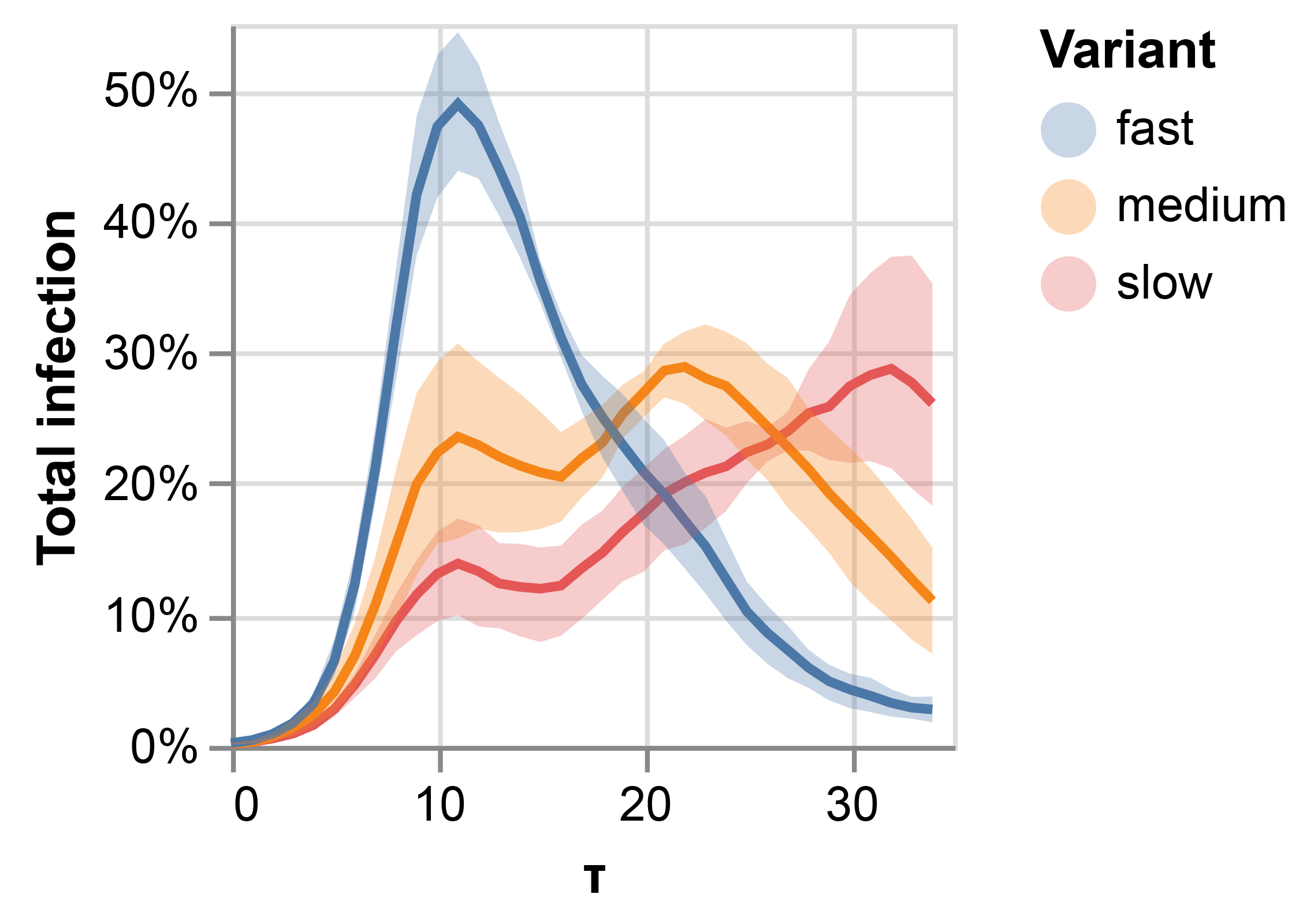} 
    \caption{No cross-immunity aggregated daily infections} 
    \label{fig:3c} 
  \end{subfigure}
   \begin{subfigure}[b]{0.5\linewidth}
    \centering
   \includegraphics[width=0.85\linewidth]{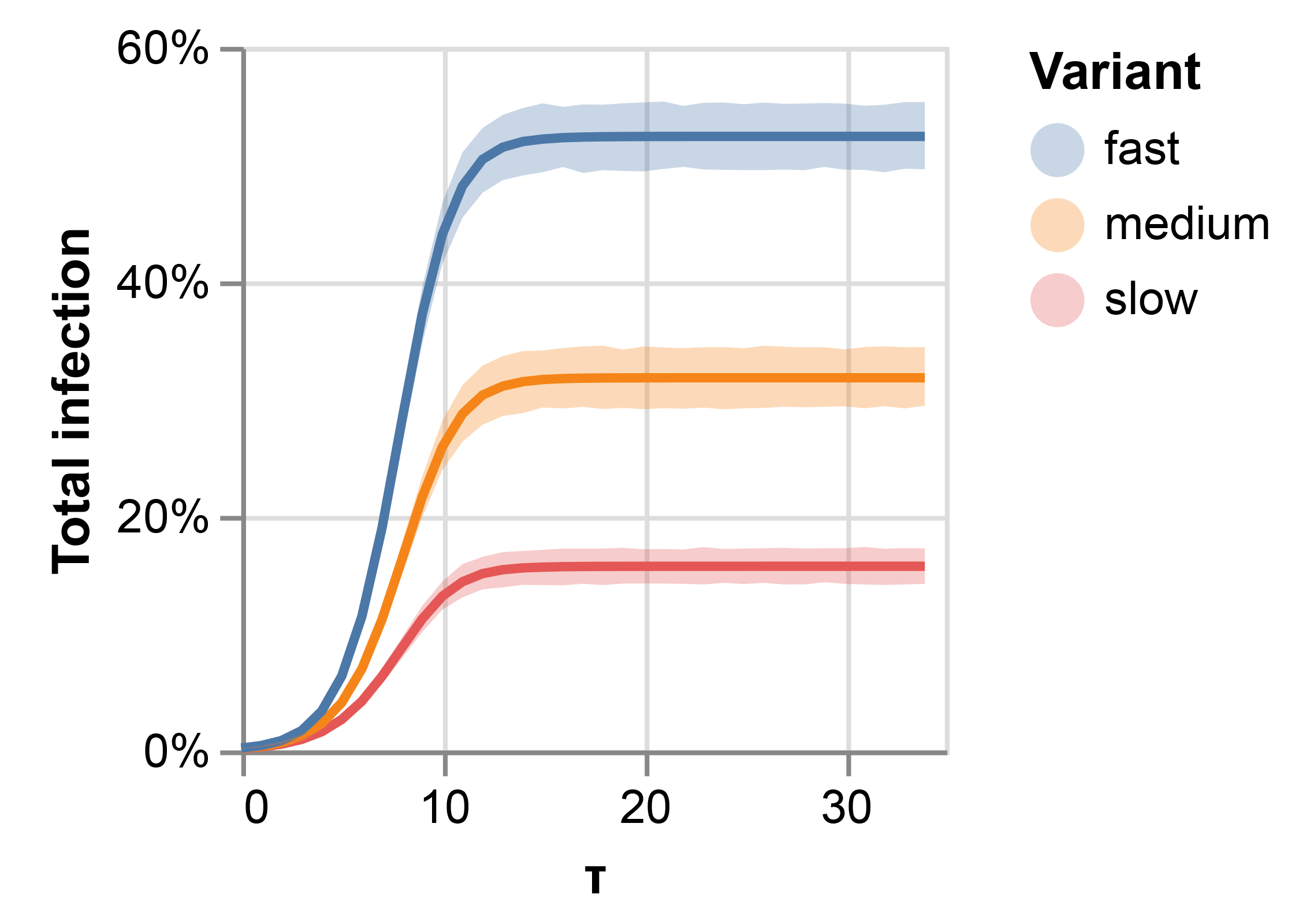} 
    \caption{Cross-immunity -  cumulative} 
    \label{fig:3b} 
    %\vspace{4ex}
  \end{subfigure} %%
  \begin{subfigure}[b]{0.5\linewidth}
    \centering
   \includegraphics[width=0.85\linewidth]{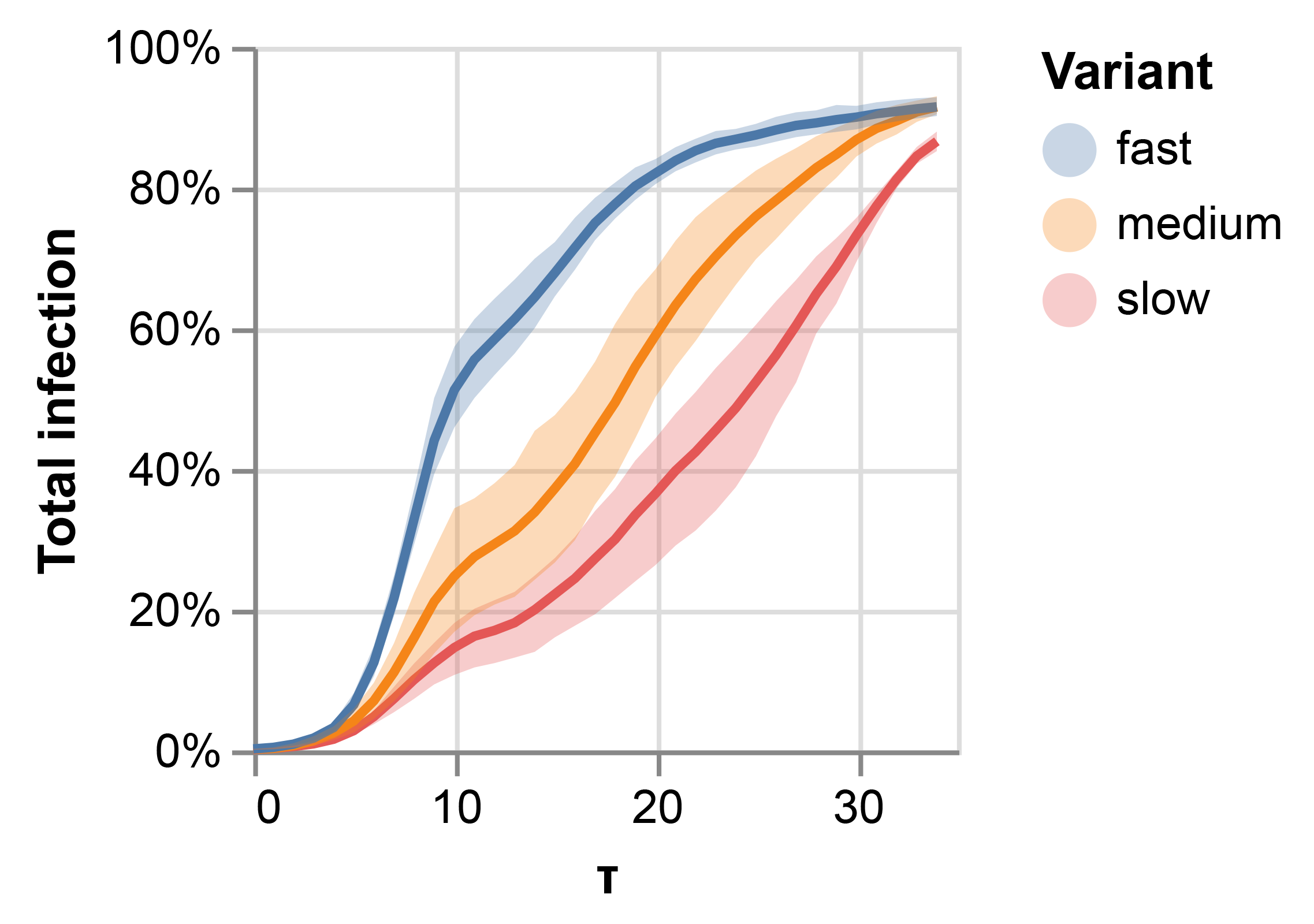} 
    \caption{No cross-immunity -  cumulative} 
     \label{fig:3d} 
  \end{subfigure} 
  \caption{Three competing variants over temporal random network. Left panels correspond to full cross-immunity conditions, right panels to no cross-immunity. }
  \label{fig:3rnd} 
\end{figure}

We continue to model the spreading of three competing variants in temporal random networks. Our modeling and considerations in Equation~\ref{eq:ie} allow for multiple variants. We then examine the competing spreading patterns under our SEIR-like model with either full cross-immunity or no cross-immunity between the variants. In the first case, infecting in any of the variants gives immunity to the other two, and in the latter, getting infected with one of the variants enables reinfection with the others. Figure~\ref{fig:3rnd} shows the results of our experiment. In Figures~\ref{fig:3a} and~\ref{fig:3b} we see the number of daily infections and the aggregated infections in the populations when there is full cross-immunity. The faster variant competes with both slower variants and infects less than 60\% of the population. The dynamics in the case of no-cross immunity, depicted in Figures~\ref{fig:3c} and~\ref{fig:3d}, are very different.  In the first wave, the variants infect at rates corresponding to their infection probability. However, once the first wave is done, the fast variant drops to the infection rate of the medium speed variant, which then ramps up and reinfects faster than before. Interestingly, the slowest variant also increases its infection rate during this second wave and maintains similar dynamics to the medium variant.  
When there is no cross-immunity between variants, competing dynamics are more complex. 

\subsection*{Real-world networks with three competing variants}
\begin{figure}[!ht] 
  \begin{subfigure}[b]{0.5\linewidth}
    \centering
   \includegraphics[width=0.9\linewidth]{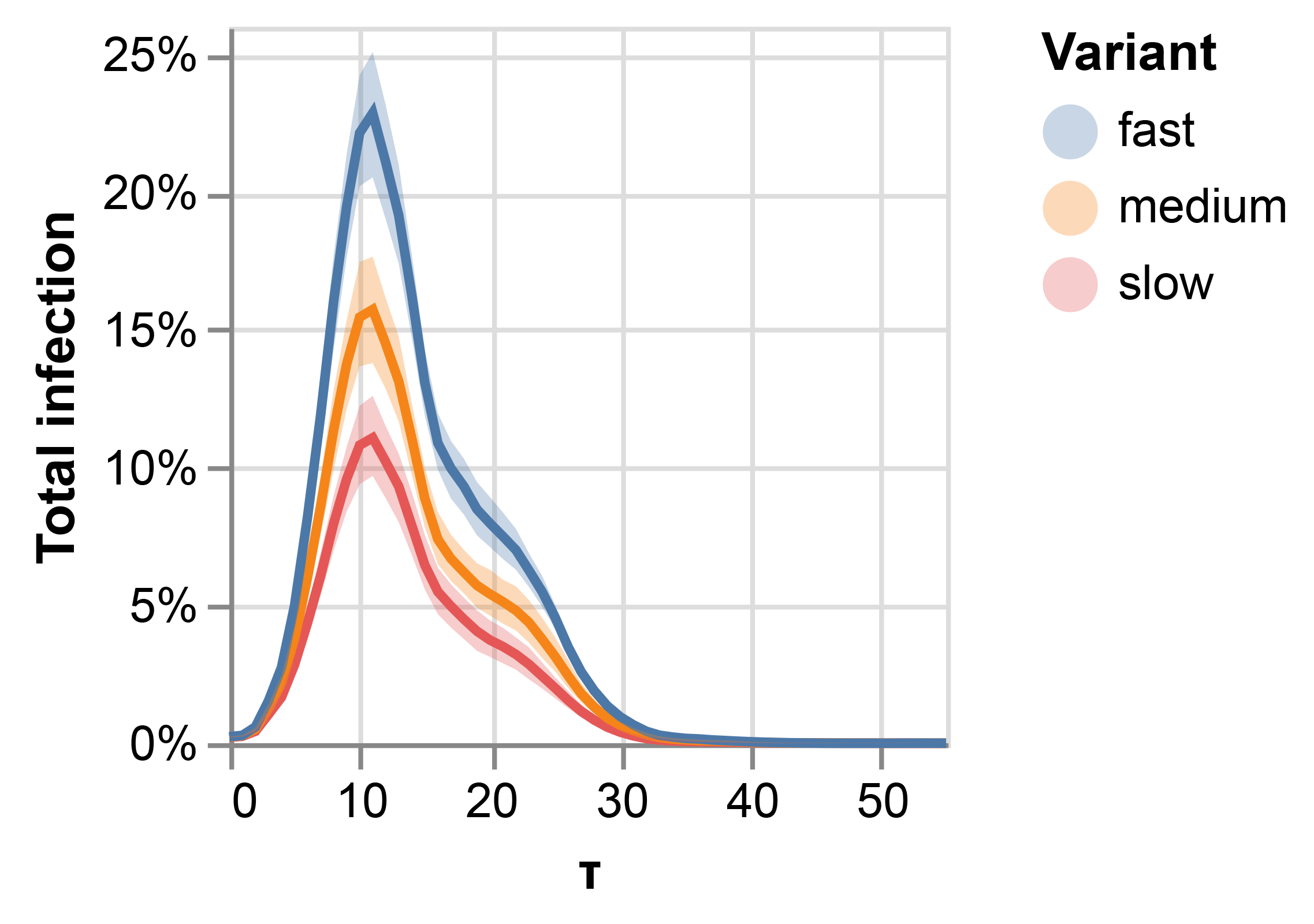} 
    \caption{} 
    \label{fig:4a} 
    %\vspace{4ex}
  \end{subfigure}%% 
  \begin{subfigure}[b]{0.5\linewidth}
    \centering
   \includegraphics[width=0.9\linewidth]{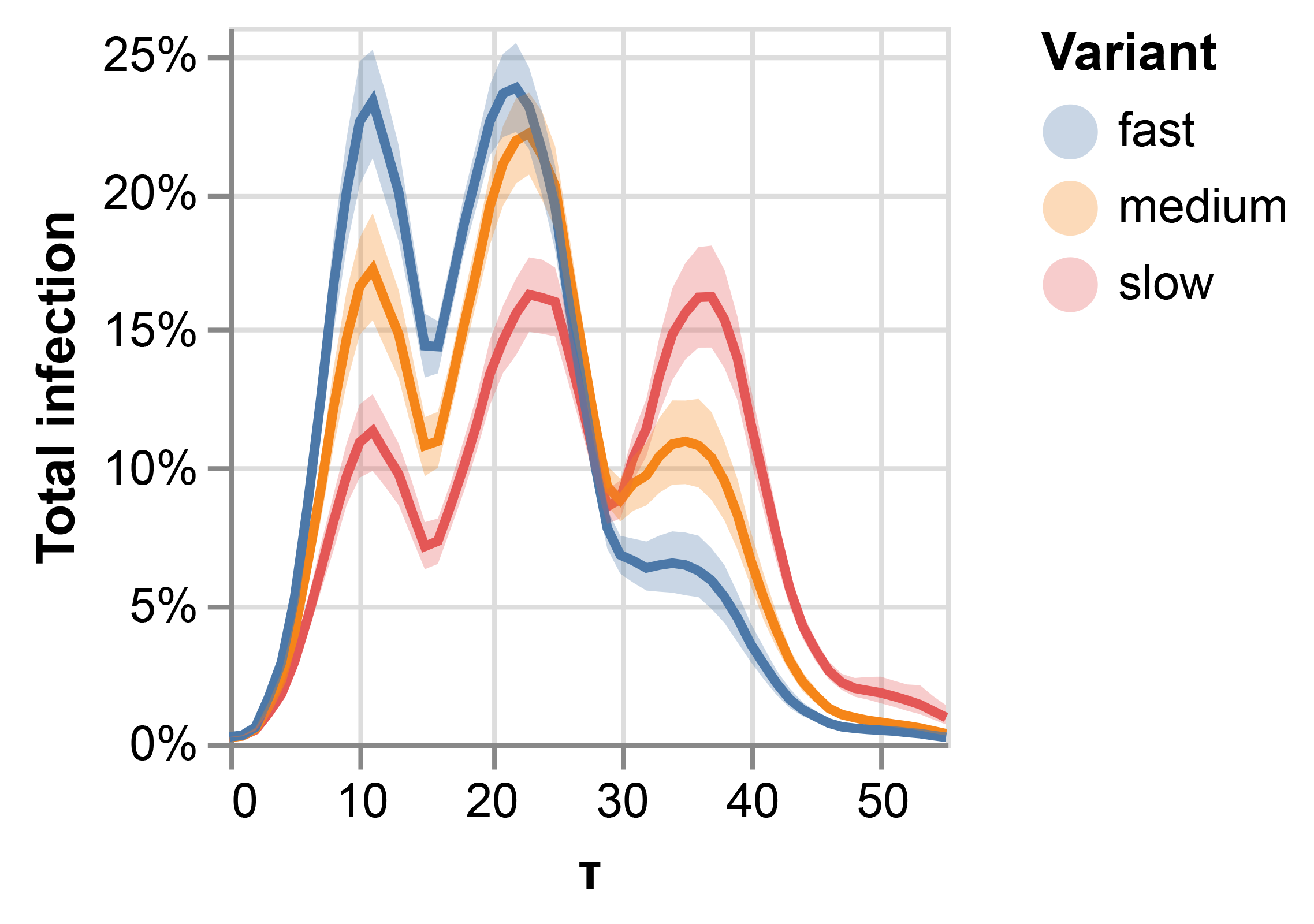} 
    \caption{} 
    \label{fig:4b} 
    %\vspace{4ex}
  \end{subfigure} 
  \begin{subfigure}[b]{0.5\linewidth}
    \centering
     \includegraphics[width=0.9\linewidth]{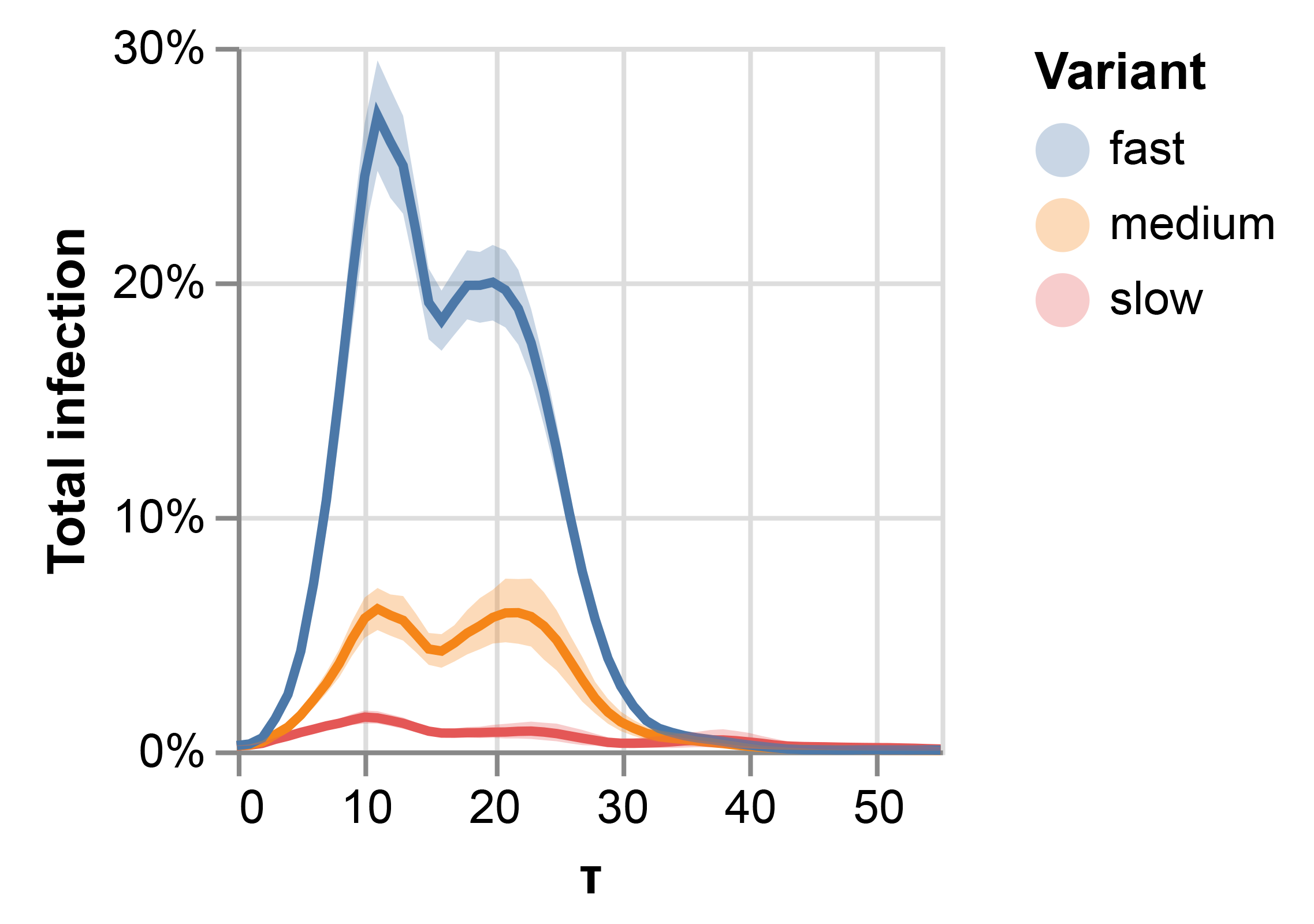} 
    \caption{} 
    \label{fig:4c} 
  \end{subfigure}%%
  \begin{subfigure}[b]{0.5\linewidth}
    \centering
    \includegraphics[width=0.9\linewidth]{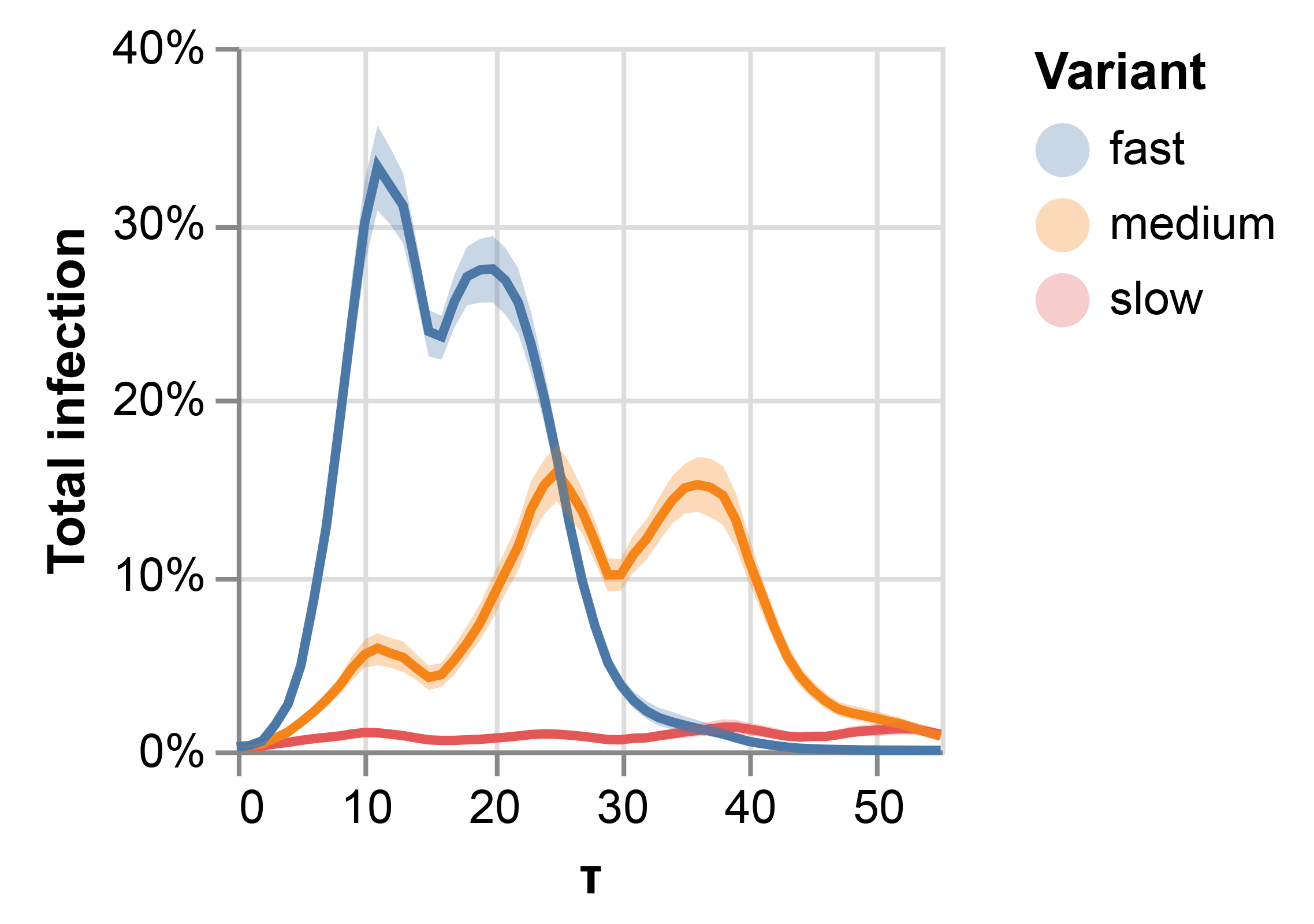} 
    \caption{} 
     \label{fig:4d} 
  \end{subfigure} 
   \begin{subfigure}[b]{0.5\linewidth}
    \centering
     \includegraphics[width=0.9\linewidth]{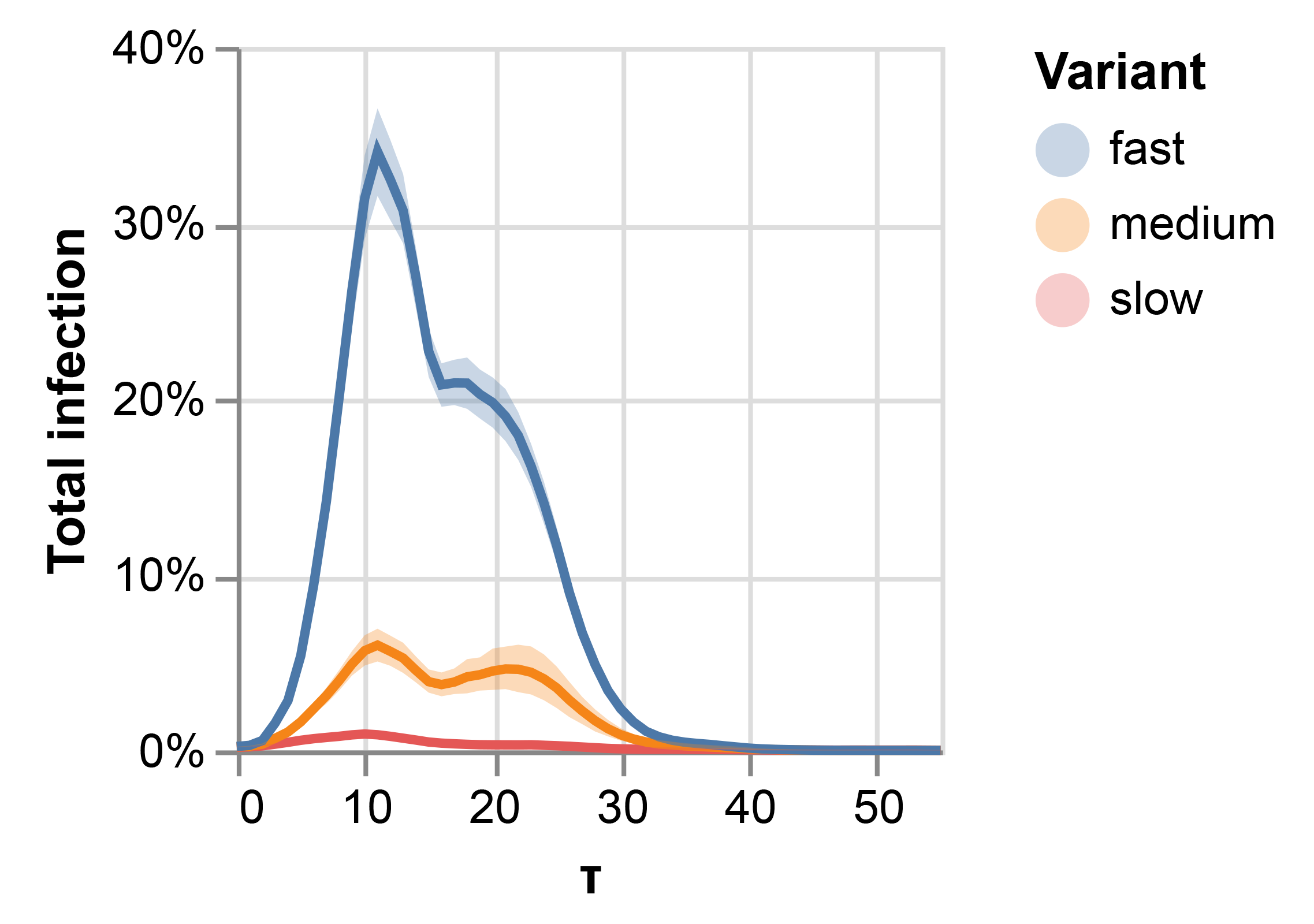} 
    \caption{} 
    \label{fig:4e} 
  \end{subfigure}%%
  \begin{subfigure}[b]{0.5\linewidth}
    \centering
    \includegraphics[width=0.9\linewidth]{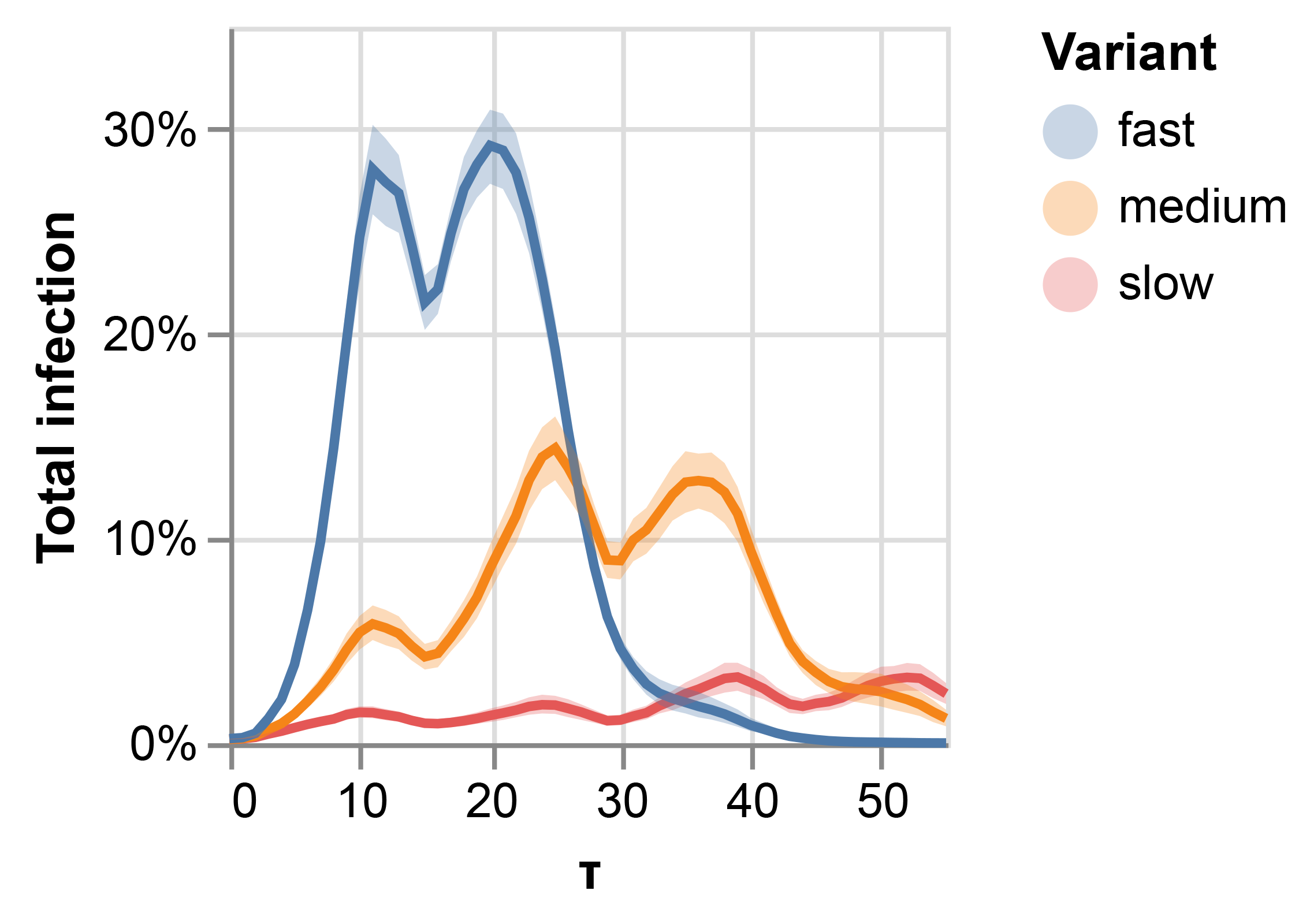} 
    \caption{} 
     \label{fig:4f} 
  \end{subfigure}  \caption{Three competing variants over real-world contact network. Left panels correspond to full cross-immunity, right panels correspond to no cross-immunity. In panels (a) and (b) we  consider different infection probabilities for the variants, but do not consider the duration of the meetings. In panels (c)-(e) we consider the duration of the encounters.  (c) and (d) depict the rate of the infection for variants that differ in the minimal time for infection. (e) and (f) show the rate of infection for variants that differ both in the minimal time for infection and infection probability. }
  \label{fig:4rwn} 
\end{figure}
We continue to evaluate the competition conditions of three different variants over the real-world temporal CNS network, described in Section~\ref{modelrw}. We model the competing variants as follows. 
\begin{description}
\item [A. ] Variants differ by the maximal probability of getting infected, as in the previous experiments. The variants were considered s.t. $$P_{\text{max}}^{\text{ fast}} = 1.2 \cdot P_{\text{max}}^{\text{ medium}} = 1.2 \cdot P_{\text{max}}^{\text{ slow}}$$
\item [B. ] The minimal duration of exposure necessary for infection, $D_{\text{min}}$, differs between variants, s.t.  $$D_{\text{min}}^{\text{ fast}} < D_{\text{min}}^{\text{ slow}}$$ Here, $$D_{\text{min}}^{\text{ fast}} = 5, D_{\text{min}}^{\text{ medium}} = 30, D_{\text{min}}^{\text{ slow}} = 55$$ As determined in Section~\ref{sec2} there is a normalizing value $D_{\text{max}}$ such that $D_{\text{min}}/D_{\text{max}}$ is the relative portion of $P_{\text{max}}$ to be used. However, in this experiment, $$P_{\text{max}}^{\text{ fast}} = P_{\text{max}}^{\text{ medium}} = P_{\text{max}}^{\text{ slow}}$$ 
\item [C. ] Variants differ by both maximal probability of getting infected and minimum duration of encounter needed for infection. That is:  $$D_{\text{min}}^{\text{ fast}} = 5, D_{\text{min}}^{\text{ medium}} = 30, D_{\text{min}}^{\text{ slow}} = 55$$ and $$P_{\text{max}}^{\text{ fast}} = 1.1\cdot P_{\text{max}}^{\text{ medium}} = 1.2 \cdot P_{\text{max}}^{\text{ slow}}$$
Also in this experiment there is a normalizing value $D_{\text{max}}$ such that $D_{\text{min}}/D_{\text{max}}$ is the relative portion of $P_{\text{max}}$ to be used.
\end{description}
Each of the experiments {\em A, B, C},  was iterated 30 times with the real-world network using the encounter-driven modeling according to the conditions of the experiment described, with an initial patient zero randomly selected. 

Figure~\ref{fig:4rwn} depicts the results of the above experiments. When the duration of the interactions is ignored, as in Figures~\ref{fig:4a} and~\ref{fig:4b}, the dynamics resemble, to some extent, those that were observed over temporal random networks. However, the dynamics differ significantly when meetings' duration is taken into account, considering that variants differ by the minimal amount of time needed for infection, as in Figures~\ref{fig:4c},~\ref{fig:4d},~\ref{fig:4e},~\ref{fig:4f}. In all these cases, the slowest variant has much fewer opportunities to infect or reinfect, causing it to disappear and leaving the two faster variants to compete. When there is full cross-immunity, the medium variant, with fewer infection opportunities than the faster one, infects at a much lower rate. However, when there is no cross-immunity, the medium-rate pathogen can create a second wave and reinfect at a higher rate than it did on the first wave.

Our results here indicate that when considering airborne diseases, it might be crucial to consider the duration of temporal meetings to consider the spreading of variants in a population.

\section*{Discussion}
\label{sec4}
Here, we presented the modeling of variants that compete over the same population with similar initial conditions over temporal networks. Previous works also considered two competing pathogens in a population. Karrer and Newman~\cite{karrer2011competing} considered competition in similar conditions, but on an aggregated network with full cross-immunity, that is, getting one of the diseases gives a subsequent immunity to both. They found that the fast pathogen is more likely to dominate the network in the majority of the regions of the phase diagram. Recently, Okake and Shudo~\cite{okabe2022spread} considered these conditions but with a lately arriving fast variant and found that the results depended on how much more infective is the added variant. They found no difference in the competition results when considering random or scale-free (BA) networks. Mann et al. ~\cite{mann2021two} showed that in this setting, clustering increases the spread of the second wave. Poletto et.al.~\cite{poletto2015characterising} considered, like us, the competition under different levels of cross-immunity, however, in a mix-homogeneous aggregated network. They found that variations in the cross-immunity level induce a transition between the presence and absence of competition. Okabe and Shudo~\cite{okabe2022spread} found that under cross-immunity conditions, faster variants dominate. We complement this finding by showing the cross-immunity conditions in which the slower variant has a second wave.   We find that when temporal paths are considered, competition conditions exist for two and three competing pathogens, and when there is no cross-immunity, there is a second wave.

This result can be considered somewhat similar to the results in~\cite{ojosnegros2012competition}, who considered competition between Colonizers, i.e., fast-spreading virulent strains and less-virulent variants that are more successful within co-infected cells. They observed a two-step dynamics of the population. Early in the infection, the population is dominated by colonizers, which later are outcompeted by competitors. We show that under certain not unlikely conditions, a slow variant might be able to wait until the fast variant has run its course and then infect the majority of the population, given there is no cross-immunity or the cross-immunity is in favor of the slower strain. 

%Yang et al. ~\cite{yang2017bi} identified that linear infection rates are overestimation of real infection rates. However, we do not consider the infection rates but the maximal probability of getting infected in a day. 
However, the competing dynamics become significantly more complex when the duration of the interactions is considered. SARS-CoV-2 variants differ in their time-to-infect~\cite{walensky2021sars,ALPERT20212595,alpert2021early,twohig2022hospital}. When we considered the duration of the interactions, the dynamics changed substantially. It is then clear that when considering airborne spreading processes, it is critical to consider the temporal ordering and the duration of the encounters. Interactions shorter than the minimal time of infection are less likely to mediate the disease, changing the progression paths. 

In summary, we model and explore competing variants using an interaction-driven model over random and real-world temporal networks. We determine the conditions of partial cross-immunity that create a second wave in which a slower  variant dominates the population. Using real-world contact data, we then show that the duration of the encounters is a fundamental modeling construct that has a significant effect on the progress of the disease and the competing dynamics of the variants.  

\section*{Methods}
\label{sec3}

 \subsection*{A temporal interaction-driven contagion model for a single variant over temporal random networks}
We explain here our model, as detailed in~\cite{abbey2022interactionbased}. We start by modeling the spread of a single variant on a temporal network. Let $G$(n,p,$\tau$) denote a temporal random network, where $\tau$ corresponds to any single time window in it.
During each time window $\tau$ we calculate for each node $i$ its probability $P_i(S \rightarrow E)$ to be exposed and infected during this time window  as the complement of probability of not being exposed to the disease in any of its encounters with infectious nodes in that time window. 
\begin{equation}
    P_i^{\tau}(S \rightarrow E)= 1-\prod^{N_i^\tau}(1- P_{\text{max}})
    \label{eq:exp} %\tag
\end{equation}

Where $N_i^\tau$ is the subset of infected nodes in time window $\tau$ that interacted with node $i$ during that time window and thus might potentially expose it to the infection, and $P_{\text{max}}$ is the probability of being infected during a maximal exposure (For example, even on relatively isolated, dense sites such as the Princess Diamond ship during the first wave of the COVID-19 pandemia, and with an air conditioning system that might well have  distributed the virus to multiple cabins, not more than 20\% of the passengers and personnel were infected. 
In other such cases, the maximal infection probability seemed to lie anywhere between 20\%  and 60\%).  

\begin{figure}[!ht]
\centering
\begin{tikzpicture}[->,>=stealth',shorten >=1pt,auto,node distance=3.2cm,
        scale = .85,transform shape][!h]
       
  \node[state] (S)  {$S$};
  \node[state] (E) [right of=S] {$E$};
  \node[state] (I) [right of=E] {$I$};
  \node[state] (R) [right of=I] {$R$};

 \path 
        (S) edge [solid] node [above]{$P_i$ (Eq. ~\ref{eq:exp})} (E)
        (S) edge [solid, loop above] node [above]{$1-P_i$ } (S)
        (E) [dashed] edge node [above] {$d \sim \tau \cdot N(5,1)$} (I)
        (I) [dashed] edge node [above] {$d \sim \tau \cdot N(10,1)$} (R);

\end{tikzpicture}
\caption{The state machine for each node in our interaction-driven SEIR-like contagion model  for random temporal networks. The transition from state $S$ to state $E$ is determined at the end of each time window $\tau$.} \label{fig:model-r}
\end{figure}
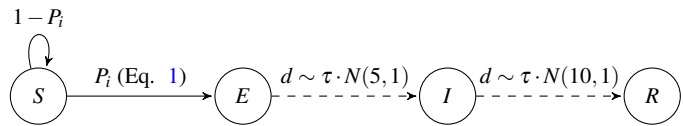

  The SEIR-like interaction-driven model, depicted in Figure~\ref{fig:model-r} is as follows. All nodes are initially in the Susceptible (\textit{S}) state. To start the iterative computation, we also assume that a small random number of nodes in the initial time window become infectious (i.e., a small number of "Patients Zero"). Susceptible nodes (state $S$) that become exposed according to Equation~\ref{eq:exp} enter the state Exposed (\textit{E}) at the end of the time window $\tau$ during which they were exposed to infectious nodes. Nodes in state Exposed become Infected after a  delay that corresponds to $ d_{E \rightarrow I}$, where $d_{E \rightarrow I} \sim \tau \cdot N(5,1)$. Infected nodes (state $I$) stay infectious for $ d_{I \rightarrow R}$, where $d_{I \rightarrow R} \sim \tau \cdot N(10,1)$. After this delay, they enter the state Recovered (\textit{R}) and are no longer infectious. The chosen numbers represent days and are inspired by some of the COVID-19 variants~\cite{hart2022generation}.

Temporal random networks follow the algorithm suggested by Zhang et al. ~\cite{zhang2017random}. Their algorithm builds a temporal random network with changing dynamics that follow a Markov process,  allowing for continuous-time network histories,  $G_{n,p,\tau}$  such that $E(G_{n,p,\tau}) \sim G_{n,p}$.  \\Defining $\lambda = \text{probability per time granule of a new edge to appear}$ and $ \mu= \text{probability per time granule of an existing edge to disappear}$, Zhang et al.~\cite{zhang2017random} show that the equilibrium probability of an edge is %\begin{equation}
\(p=\frac{\lambda}{\lambda+\mu}\). 

%\end{equation} 
 Our implementation, a Python package that we refer to as RandomDynamicGraph (RDG), (\label{github_lib}\url{https://github.com/ScanLab-ossi/DynamicRandomGraphs}) described in~\cite{abbey2022interactionbased}, generates large-scale dynamic random graphs according to a defined density. We chose $\lambda, \mu$ in a manner that allows for a stationary density for all $G_{n,p,\tau}$, where $\tau \in [1..10000]$ and $n=1000$.  
To ensure that no anomalies in one graph created an unusual effect, we used a mix of randomly generated graphs. We then conducted a Two-sample Kolmogorov–Smirnov test on each pair of graphs and determined that for each couple, we cannot reject the null hypothesis that both were drawn from the same probability distribution, with 95\% significance~\cite{miller2020size}.

\subsection*{Modeling multiple concurrent infections}
We model the various variants by a per-variant  probability, denoting the maximal global probability of being infected by the variant, $P_{\text{max}}^j$, where $j$ denotes the variant.  Hence, a more contagious variant would correspond to a higher $P_{\text{max}}$. Thus, given variants $j,k$, then, without loss of generality, if variant $j$ is more contagious than variant $k$, then $P_{\text{max}}^{j} > P_{\text{max}}^{k}$.

We model here the probability to be infected during a day when $m$ variants exist in the network, as follows. Let $P^{k}_i(S \rightarrow E)$ be the probability of node $i$ to be exposed during some given time window to variant $k, k \in [1..m]$. $P^{k}_i(S \rightarrow E)$ is calculated as is defined in Equation~\ref{eq:exp}. 
Then, we calculate the probability of node $i$ to not  be exposed to any of the $m$ variants during that time window, $P_i(S \rightarrow S)$
using the inclusion-exclusion principle. Given that for any $k$, $P^{k}_i(S \rightarrow E)$ is independent, i.e., each viral strain is operating independently of other strains, the probability of {\em not} getting infected with any variant  is as follows. Let $E^{i}_j$ be the event of node $i$ being exposed by the end of the day to variant $j$, the probability that node $i$ was not exposed to any variant at the end of the day, is computed as follows:
\begin{equation}
   P_i(S \rightarrow S) = 1 -  \sum_{k=1}^m \left( (-1)^{k+1}  \sum_{1 \leq \cdots < k \leq m} \mid E^i_{1} \cap \cdots \cap E^i_{k} \mid \right)  
   \label{eq:ie}
\end{equation}

For example, for two variants $j,k$, we calculate the probability of node $i$ not being exposed to any of them during that day as follows, thus determining the probability of being infected in any of them. 
\[P_i(S \rightarrow S)  = 1 - (P_{\textit{exposed}}^{j} + P_{\textit{exposed}}^{k} - P_{\textit{exposed}}^{j}\cdot P_{\textit{exposed}}^{k}) \]

%We assume that there are $m$ variants. The probability to be infected in a time-window is calculated separately for each variants, according to Equation~\ref{eq:exp}. 
%Then, given the probability of node $i$ as calculated in Equation~\ref{eq:exp} to be exposed to variant $j_k, k \in [1..m]$ is $P^{j_k}_i(S \rightarrow E)$, we calculate the probability of node $i$ to not be exposed to any of the variants during time window $\tau$
%using the inclusion-exclusion principle, as follows: 

Given that we determined that node $i$ was exposed, we make a statistical choice of the variant according to the relative probabilities.

\subsubsection*{Variants cross-immunity probability}\label{sec2}

Each variant has a cross-immunity probability $\chi_{j \rightarrow k}$, denoting the probability of a person that is recovering from variant $j$ to be immune to variant $k$. Hence, $1-\chi_{j \rightarrow k}$ is the breakthrough probability of people recovering from variant $j$ to get infected by variant $k$.
$\chi_{j \rightarrow k} \in [0,1]$.

\subsection*{Modeling variants competition over real-life encounters}\label{modelrw}
Real-world encounters data was collected in the Copenhagen Networks Study (CNS)~\cite{stopczynski2014measuring,sapiezynski2019interaction}. It contains the proximity information of over 700 students for 28 days. We model the CNS social network of interactions  $\Gamma$ as a sequence of $T$ consecutive undirected weighted temporal graphs $\{G_\tau \in \Gamma, \tau \in T\}$ where each temporal snapshot graph $G_\tau=(V_\tau,E_\tau)$ denotes the subset of interacting nodes $V_\tau$ during the $\tau$ temporal window and the weighted edges $E_\tau$ the interactions during this time. 
Further, we use the network with half-density, i.e., each real day is split into two days. This is done since the original proximity network is very dense~\cite{Genois2018,abbey2022interactionbased}. %and in very dense networks, variants with various contagious levels spread at a similar rate~\cite{abbey2022interactionbased}. 

The CNS data contains meetings' duration information, allowing for an additional method of modeling the variants. In this method, a variant takes a minimal amount of time to infect, termed $D_{\text{min}}$, such that $D_{\text{min}}^{\text{ fast}} < D_{\text{min}}^{\text{ slow}}$. There is a normalizing value $D_{\text{max}}$ such that $D_{\text{min}}/D_{\text{max}}$ is the relative portion of $P_{\text{max}}$ to be used. The contagious part of the interaction-driven model is then defined by the complement of the probability of not being infected by any of the encounters that are longer than $D_{\text{min}}$ of that variant in a day~\cite{abbey2022interactionbased}. 

\bibliography{competing}

%\noindent LaTeX formats citations and references automatically using the bibliography records in your .bib file, which you can edit via the project menu. Use the cite command for an inline citation, e.g.  \cite{Hao:gidmaps:2014}.

%For data citations of datasets uploaded to e.g. \emph{figshare}, please use the \verb|howpublished| option in the bib entry to specify the platform and the link, as in the \verb|Hao:gidmaps:2014| example in the sample bibliography file.

\section*{Acknowledgements}
Alex Abbey and Osnat Mokryn were partially supported by the Israeli Science Foundation grant 328/17. 

\section*{Author contributions statement}
OM, YS, AA designed the experiments; AA wrote the code and performed all the experiments; All authors analyzed the results; OM wrote the paper; All authors reviewed the manuscript.
\section*{Additional information}

\textbf{The authors declare no competing interests.}  \\
\textbf{Code availability:} All code and data used in this research are freely available: \url{https://github.com/ScanLab-ossi/covid-simulation}.
%The corresponding author is responsible for submitting a \href{http://www.nature.com/srep/policies/index.html#competing}{competing interests statement} on behalf of all authors of the paper. This statement must be included in the submitted article file.

\end{document}